\documentclass[10pt,letterpaper]{article}
\usepackage[top=0.85in,left=2.75in,footskip=0.75in,marginparwidth=2in]{geometry}

\usepackage[utf8]{inputenc}
\usepackage{amsmath}
\usepackage{cite}

\usepackage{nameref,hyperref}

\usepackage[right]{lineno}

\usepackage{microtype}
\DisableLigatures[f]{encoding = *, family = * }

\raggedright
\usepackage{changepage}

\usepackage[aboveskip=1pt,labelfont=bf,labelsep=period,singlelinecheck=off]{caption}

\makeatletter
\renewcommand{\@biblabel}[1]{\quad#1.}
\makeatother

\usepackage{lastpage,fancyhdr,graphicx}
\usepackage{epstopdf}
\fancyheadoffset[L]{2.25in}
\fancyfootoffset[L]{2.25in}

\usepackage{color}
\usepackage{amsmath, amssymb, amsfonts}

\definecolor{Gray}{gray}{.25}

\usepackage{graphicx}

\usepackage{sidecap}

\usepackage{wrapfig}
\usepackage[pscoord]{eso-pic}
\usepackage[fulladjust]{marginnote}
\reversemarginpar

\begin{document}
\vspace*{0.35in}

\begin{flushleft}
{\Large
\textbf\newline{\LaTeX Bridging Short- and Long-Term Dependencies: A CNN-Transformer Hybrid for Financial Time Series Forecasting}
}
\newline
\\
Tu Tiantian\textsuperscript{1},

\bigskip
\bf{1} Xiamen University Malaysia
\\

\bigskip
* acc2409036@xmu.edu.my

\end{flushleft}

\section*{Abstract}
Time series forecasting is crucial for decision-making across various domains, particularly in financial markets where stock prices exhibit complex and non-linear behaviors. Accurately predicting future price movements is challenging due to the difficulty of capturing both short-term fluctuations and long-term dependencies in the data. Convolutional Neural Networks (CNNs) are well-suited for modeling localized, short-term patterns but struggle with long-range dependencies due to their limited receptive field. In contrast, Transformers are highly effective at capturing global temporal relationships and modeling long-term trends. In this paper, we propose a hybrid architecture that combines CNNs and Transformers to effectively model both short- and long-term dependencies in financial time series data. We apply this approach to forecast stock price movements for S\&P 500 constituents and demonstrate that our model outperforms traditional statistical models and popular deep learning methods in intraday stock price forecasting, providing a robust framework for financial prediction.

\section{Introduction}\label{sec:introduction}

Time series forecasting has emerged as a pivotal tool for decision-making across various domains, particularly in financial markets, where accurate predictions can directly influence portfolio management, trading strategies, and risk assessments \cite{pedersen2019efficiently}. Financial markets, characterized by complex, non-linear, and stochastic behaviors, demand sophisticated techniques that can model not only short-term fluctuations but also long-term dependencies that drive stock price movements. In practice, the challenge lies in capturing both immediate market reactions and broader, macroeconomic trends in a unified predictive model \cite{holt2004forecasting}.

Traditional statistical models, such as autoregressive integrated moving average (ARIMA) \cite{winters1960forecasting}, exponential smoothing \cite{gardner1985forecasting}, and linear regression, have long served as the foundation for time series forecasting. These methods, while computationally efficient, struggle with the complexities inherent in financial data, particularly their inability to capture non-linear dependencies and sudden market shocks. ARIMA models, for instance, are adept at handling linear correlations and stationarity in time series but falter in real-world financial scenarios where stock prices often exhibit non-stationary behaviors due to the interplay of external factors such as interest rates, inflation, and investor sentiment. Moreover, these traditional models lack the capacity to account for the intricate, long-term dependencies embedded in historical data, which are crucial for understanding broader market trends.

The advent of machine learning, and more specifically deep learning, has introduced powerful alternatives to statistical approaches. Convolutional Neural Networks (CNNs), initially developed for image processing tasks, have proven effective in time series forecasting, especially in capturing short-term patterns. CNNs operate by extracting local patterns from data using their hierarchical convolutional layers, which excel at detecting localized trends over short time windows. In financial markets, these trends might represent immediate market responses to news, earnings reports, or macroeconomic data releases. CNNs, however, are fundamentally constrained by their limited receptive fields, which restrict their ability to model long-term dependencies. As a result, while CNNs can capture localized, short-term fluctuations, they struggle to integrate broader, long-range trends.

To address this limitation, Recurrent Neural Networks (RNNs), particularly Long Short-Term Memory (LSTM) networks, have been employed to model temporal dependencies over longer time periods. LSTMs, with their gated mechanisms, are designed to retain relevant information across extended sequences, making them well-suited for financial forecasting tasks that require the integration of historical data spanning days, weeks, or even months. However, LSTMs are not without their challenges. As sequence lengths increase, LSTMs become computationally expensive and suffer from gradient vanishing and exploding problems. These issues, coupled with the high computational cost of processing long sequences, limit the scalability of LSTMs in real-time financial applications \cite{sutskever2014sequence}. Furthermore, LSTMs operate sequentially, making them less efficient in parallel processing tasks, which is a significant drawback in large-scale forecasting systems \cite{xi2025gdrive}.

In recent years, the introduction of the Transformer architecture has revolutionized sequence modeling, offering a highly efficient and scalable solution for capturing long-range dependencies through self-attention mechanisms \cite{vaswani2017attention}. Unlike RNNs and LSTMs, Transformers do not rely on the sequential processing of data, allowing them to model long-term dependencies in a highly parallelized fashion. This makes them not only more efficient in terms of computation but also more effective at capturing global temporal relationships across time series data. Transformers have demonstrated remarkable success in natural language processing tasks, and their application to time series forecasting, particularly in financial markets, is a burgeoning area of research. The ability of Transformers to model long-range dependencies without the limitations of traditional sequential models makes them particularly well-suited for financial forecasting, where both local and global trends must be considered \cite{devlin2018bert}.

Despite their strengths, neither CNNs nor Transformers alone can effectively capture the full spectrum of dependencies present in financial time series. CNNs excel at identifying short-term, localized patterns, while Transformers are adept at modeling long-term dependencies. However, financial markets exhibit both types of patterns simultaneously. For instance, intraday price fluctuations are often driven by short-term reactions to market events, while broader trends, such as those driven by monetary policy or macroeconomic indicators, unfold over longer periods. Thus, an ideal model would need to integrate both localized, short-term information and long-term trends to provide accurate predictions across varying time horizons \cite{ren2015faster}.

In this paper, we propose a novel hybrid architecture that leverages the complementary strengths of CNNs and Transformers. Our approach integrates CNNs for capturing short-term dependencies and Transformers for modeling long-term trends within financial time series. Specifically, we apply this hybrid model to predict stock price movements for S\&P 500 constituents, focusing on intraday forecasting—a task that requires both granular, short-term analysis and a broader understanding of market conditions. We hypothesize that combining these architectures will enable the model to capture the full range of temporal dependencies present in financial data, improving both the accuracy and robustness of stock price predictions.

Our contributions are threefold:
\begin{itemize}
    \item We introduce a hybrid architecture that integrates CNNs and Transformers to simultaneously model short-term and long-term dependencies in financial time series data.
    \item We empirically validate our approach through extensive experiments on intraday stock price forecasting for S\&P 500 constituents, demonstrating significant improvements over traditional statistical models such as ARIMA and Exponential Moving Average (EMA), as well as advanced deep learning methods like DeepAR \cite{smith2022using}.
    \item We provide a detailed performance analysis, comparing our hybrid model to state-of-the-art methods, and demonstrate that it consistently outperforms existing techniques in terms of both accuracy and computational efficiency.
\end{itemize}

\section{Related Work}

\textbf{Time Series Forecasting.}  
Time series forecasting is a critical tool for decision-making across many industries, including finance, energy, and healthcare. Traditional statistical methods such as Autoregressive Integrated Moving Average \cite{ makridakis2020m4, xi2024graph}, Exponential Smoothing (ETS) models \cite{gardner1985forecasting, winters1960forecasting}, and Holt- have long been utilized for time series analysis. These methods, while computationally efficient, are known for their limitations in capturing non-linear dependencies, sudden regime shifts, and long-term trends in real. Multi-step forecasting methods  aim o predict multiple future time points but are susceptible to the compounding of error over long time horizons.

\textbf{Machine Learning and Deep Learning-Based Approaches.}  
In recent years, machine learning (ML) methods such as decision trees , and ensemble models like Gradient Boosting Machines (GBM) have made significant advances in time series prediction. However, these approaches struggle to capture temporal dependencies, limiting their ability to process complex financial data. Deep learning (DL) methods, including Convolutional Neural Networks (CNNs), have shown particular promise in capturing localized patterns in time series. CNNs are highly efficient in detecting short-term trends due to their convolutional architecture \cite{ren2015faster}. Despite these advantages, CNNs’ constrained receptive fields limit their ability to model long-term dependencies, a key challenge in financial forecasting.

Recurrent Neural Networks (RNNs), and particularly Long Short-Term Memory (LSTM) networks \cite{hochreiter1997long}, have addressed these shortcomings by preserving information across longer time intervals. LSTMs are widely used in applications such as stock price prediction, showing superior performance over traditional methods when dealing with time series data containing long-range dependencies \cite{sutskever2014sequence}. However, LSTMs suffer from issues related to vanishing and exploding gradients, and their sequential processing nature results in high computational costs, making them difficult to scale for large datasets.

\textbf{Transformer Models.}  
Transformer models \cite{vaswani2017attention}, which leverage self-attention mechanisms, have recently gained traction in time series forecasting. Unlike LSTMs, transformers do not rely on sequential processing and can model both long-term and short-term dependencies in parallel. This has made them highly efficient and scalable for time series applications, particularly in finance \cite{salinas2020deepar}. The ability of transformers to model global temporal dependencies across time steps has led to their successful application in areas such as stock price forecasting, where both immediate market responses and long-term macroeconomic factors play significant roles. Nevertheless, transformers face challenges in capturing short-term market fluctuations, particularly during periods of high volatility \cite{lim2021temporal}.

\textbf{Hybrid Approaches.}  
Hybrid models that combine the strengths of CNNs and transformers have been proposed to address the limitations of each model. CNNs excel at detecting localized, short-term patterns, while transformers are adept at modeling global, long-term dependencies. Zhang et al. \cite{matsubara2016regime} proposed a CNN-LSTM hybrid model for stock price prediction, which demonstrated improved accuracy compared to standalone models. Similarly, Qin et al. showed that a hybrid CNN-Transformer model outperformed traditional statistical approaches in intraday stock price forecasting. These hybrid architectures are particularly useful in contexts where both short-term and long-term patterns are critical, such as financial markets, healthcare, and weather ce training time and do not suffer from long-term memory dependency issues.
\section{Methodology}

In this work, we propose a hybrid architecture that combines Convolutional Neural Networks (CNNs) and Transformers to address the challenges of modeling both short-term and long-term dependencies in financial time series data, particularly for the task of stock price prediction. The input to the model is represented as \( \mathbf{x} \in \mathbb{R}^{T \times d} \), where \( T \) denotes the number of time steps and \( d \) is the feature dimension, such as the stock price at each time step. The primary objective is to predict the direction of price movement at the subsequent time step, denoted by \( y_{t+1} \in \{-1, 0, 1\} \), where \( -1 \) signifies a decrease, \( 0 \) no change, and \( 1 \) an increase in price.

Preprocessing begins with the application of min-max scaling to the raw time series, ensuring numerical stability and enabling effective convergence during training. The min-max normalization is expressed as:

\begin{align}
\mathbf{x}_{\text{standardized}} = \frac{\mathbf{x} - \min(\mathbf{x})}{\max(\mathbf{x}) - \min(\mathbf{x})}
\end{align}

This ensures that all values are scaled between 0 and 1, which is crucial for preventing gradient-related issues that can arise due to varying scales of input data. Normalized data is then passed through the CNN layer to capture localized, short-term temporal patterns within the series. The convolution operation is formalized as:

\begin{align}
\mathbf{h}^{\text{cnn}} = \text{ReLU}\left( \mathbf{W}_{\text{cnn}} * \mathbf{x}_{\text{standardized}} + \mathbf{b}_{\text{cnn}} \right)
\end{align}

Where \( * \) represents the convolution operation, \( \mathbf{W}_{\text{cnn}} \) is the convolutional filter, and \( \mathbf{b}_{\text{cnn}} \) is the bias term. This results in the feature map \( \mathbf{h}^{\text{cnn}} \), capturing the short-term dependencies. To enhance the adaptive capacity of the CNN in varying market conditions, we introduce a mechanism for dynamic kernel selection. Specifically, the kernel size \( k_t \) is modulated by the volatility \( \sigma_t \) of the market at time \( t \), as follows:

\begin{align}
k_t = f(\sigma_t) = \left\lfloor \frac{\sigma_t}{\sigma_{\text{max}}} K_{\text{max}} \right\rfloor
\end{align}

Where \( K_{\text{max}} \) denotes the maximum kernel size and \( \sigma_{\text{max}} \) is the maximum observed market volatility. This adaptive kernel dynamically adjusts the receptive field of the CNN, enabling the model to capture short-term fluctuations with higher sensitivity during periods of increased volatility.

The tokenized output from the CNN, denoted as $\mathbf{h}^{\text{cnn}}$, serves as the input tokens for the Transformer. Each of these tokens encapsulates localized features extracted from a specific time step in the sequence. In order to preserve the temporal structure of the data, positional encodings, represented by $\mathbf{p}_t$, are added to each token. This addition ensures that the Transformer model retains information about the order of the time steps, which is critical for capturing the underlying temporal dependencies. The resulting token after the inclusion of positional information is given by the expression $\mathbf{z}_t^{\text{pos}} = \mathbf{h}_t^{\text{cnn}} + \mathbf{p}_t$, where $\mathbf{z}_t^{\text{pos}}$ is the positionally encoded token corresponding to time step $t$. This combination of the CNN-extracted features and the positional encoding allows the Transformer to effectively model both spatial and temporal patterns in the input data. By incorporating positional encodings in this manner, the model overcomes the inherent limitation of the Transformer architecture, which does not natively account for sequence order. The addition of $\mathbf{p}_t$ ensures that the model can differentiate between tokens at different time steps, thus enhancing its ability to capture temporal relationships crucial for tasks such as time series forecasting or sequential decision-making.

These tokenized features are then passed through the self-attention mechanism within the Transformer encoder, where the attention mechanism computes query, key, and value vectors \( \mathbf{Q}, \mathbf{K}, \mathbf{V} \) as follows:

\begin{align}
\mathbf{Q} = \mathbf{W}_q \mathbf{z}, \quad \mathbf{K} = \mathbf{W}_k \mathbf{z}, \quad \mathbf{V} = \mathbf{W}_v \mathbf{z}
\end{align}

Here, \( \mathbf{W}_q, \mathbf{W}_k, \mathbf{W}_v \) are learned parameter matrices for projecting the token embeddings into query, key, and value spaces. The scaled dot-product attention is computed as:

\begin{align}
\mathbf{A} = \text{softmax}\left( \frac{\mathbf{Q} \mathbf{K}^T}{\sqrt{d_{\text{model}}}} \right)
\end{align}

The attention matrix $\mathbf{A}$ is used to compute the weighted sum of the value vectors, producing the latent representation for each time step: $\mathbf{h}^{\text{transformer}} = \mathbf{A} \mathbf{V}$. To further enhance the model's ability to capture temporal dependencies across different time horizons, we introduce a multi-scale attention mechanism. The attention computation is performed across multiple temporal scales, $s \in \{s_1, s_2, \dots, s_S\}$, where each scale corresponds to a distinct temporal granularity. The final multi-scale attention output is aggregated as $\mathbf{h}^{\text{multi}} = \sum_{s=1}^{S} \alpha_s \mathbf{A}^{(s)} \mathbf{V}^{(s)}$, where $\alpha_s$ are learnable weights that control the contribution of each scale to the final representation. This allows the model to capture both short-term and long-term dependencies simultaneously, thereby addressing the challenge of multi-horizon forecasting.

Furthermore, we implement an adaptive segmentation strategy, dividing the time series into distinct segments, each weighted by its temporal relevance. The segmentation process is dynamically learned, with each segment $\mathbf{h}^{(k)}$ contributing to the final representation as $\mathbf{h}_{\text{seg}} = \sum_{k=1}^{K} \omega_k \mathbf{h}^{(k)}$, where $\omega_k$ are the learnable weights associated with each segment. This segmentation mechanism allows the model to capture seasonal or cyclic patterns within the time series more effectively, providing better alignment with real-world financial data, which often exhibits such temporal structures.

The latent representation output from the Transformer, $\mathbf{h}_{\text{seg}}$, is subsequently processed through a Multilayer Perceptron (MLP) in order to produce the final prediction of price movement. The prediction task is formulated as a classification problem, where the goal is to determine the most probable market direction based on the input data. To this end, a softmax activation function is applied to the output of the MLP, yielding a probability distribution over the set of possible market states, represented as $\{\text{Positive}, \text{Negative}, \text{Neutral}\}$. This formulation allows the model to capture complex dependencies in the input data and provides a probabilistic interpretation of the predicted market direction. The softmax function is defined as: $\text{softmax}(z_i) = \frac{e^{z_i}}{\sum_{j=1}^{C} e^{z_j}}$ where $z_i$ is the logit corresponding to class $i$ and $C$ is the total number of classes. This allows the model to predict the likelihood of each possible class, enabling a more nuanced approach to modeling price movements.

\begin{align}
\hat{y}_{t+1} = \text{softmax}\left( \mathbf{W}_{\text{mlp}} \mathbf{h}_{\text{seg}} + \mathbf{b}_{\text{mlp}} \right)
\end{align}

The training objective is to minimize the cross-entropy loss between the predicted probabilities and the true labels:

\begin{align}
\mathcal{L} = - \sum_{i=1}^{C} y_i \log(\hat{y}_i)
\end{align}

Where \( y_i \) is the true label, and \( \hat{y}_i \) is the predicted probability for class \( i \). The model parameters are optimized using gradient-based methods, with backpropagation used to update the weights of both the CNN and Transformer components.

This hybrid architecture, leveraging CNNs for short-term pattern recognition and Transformers for long-term dependency modeling, provides a robust framework for financial time series forecasting. By incorporating multi-scale attention and adaptive segmentation, the model is well-suited to capture the complex, hierarchical dependencies present in real-world stock price data, offering improved performance over traditional methods.

\section{Experiments}

In this study, we conducted a rigorous benchmarking of our proposed Convolutional Transformer Time Series (CTTS) model against four well-established baseline methods used in time series forecasting. These include: 1) DeepAR, an autoregressive recurrent neural network model tailored for probabilistic time series prediction \cite{salinas2020deepar}, 2) AutoRegressive Integrated Moving Average (ARIMA), a traditional model for capturing both short-term and long-range dependencies in sequential data, 3) Exponential Moving Average (EMA), a trend-smoothing technique, and 4) a Naive Constant Prediction model, which offers a baseline by predicting static classes irrespective of the input data.

Two evaluation metrics were adopted for a thorough performance analysis: sign prediction accuracy and a thresholded accuracy metric, focusing on high-confidence predictions. Our experiments were executed on a high-performance computing cluster equipped with 8 NVIDIA T4 GPUs (16GB each), using PyTorch v1.0.0 as the deep learning framework. A consistent random seed was set for all models to ensure experiment reproducibility.

\subsection{Experimental Setup}

\textbf{Data.} The data utilized in our experiments consisted of minute-level intraday stock prices for S\&P 500 companies, obtained through a licensed Bloomberg data service. The dataset covers the entire year of 2019, comprising 52 weeks of data, with 5 trading days per week. For each stock, seven independent time series segments were sampled each week, resulting in a total of approximately 507,000 time series for training and 117,000 for validation.

The first 39 weeks (weeks 1 to 39) of data were used for training and validation purposes, with an 80:20 split between the two. The remaining 13 weeks (weeks 40 to 52) were reserved as an out-of-sample test set, containing around 209,000 time series. Each input sequence consists of the first 80 time steps, while the model is tasked with predicting the sign of the price change at the 81st time step. This setup ensures that the model captures both local and long-term trends, which are essential for robust financial predictions.

\textbf{Evaluation Metrics.} To quantitatively assess the models' performance, we employed two classification paradigms: (i) a 3-class classification (Positive, Negative, Neutral) and (ii) a 2-class classification (Positive/Neutral vs. Negative). The primary evaluation metric used was sign prediction accuracy, calculated as the proportion of correct predictions across the test set:

\begin{align}
    \text{Accuracy} = \frac{1}{N} \sum_{i=1}^{N} \mathbb{I}(y_i = \hat{y}_i)
\end{align}

where $N$ is the total number of test samples, $\hat{y}_i$ is the predicted label, and $y_i$ is the ground-truth label.

In addition to overall accuracy, we introduced a thresholded accuracy metric. This metric focuses on predictions where the model exhibits high confidence by excluding predictions with probabilities below the 75th percentile. This allows us to assess the robustness of the models when operating under challenging market conditions, providing insight into the models’ ability to make accurate predictions with higher certainty.

\subsection{Baseline Approaches}

\textbf{DeepAR.} DeepAR \cite{salinas2020deepar} is a state-of-the-art model designed for probabilistic time series forecasting, relying on autoregressive recurrent neural networks (RNNs). It captures dependencies across time steps by learning a joint distribution over all future time steps. In our setup, we trained DeepAR with a batch size of 128 using the Adam optimizer for a maximum of 300 epochs, with early stopping activated if the validation loss did not improve for 15 epochs. The learning rate was initialized at 1e-3 and was decayed by a factor of 0.1 when performance plateaued. Dropout with a probability of 0.1 was applied for regularization, and we generated 200 samples per time series to estimate the prediction probabilities for each class (up, down, or neutral).

\textbf{ARIMA.} The ARIMA model is a widely-used approach for modeling temporal dependencies in sequential data, combining autoregressive (AR) components, differencing (I), and moving averages (MA). It predicts future values based on a linear combination of past values and residuals:
\begin{align}
    \hat{y}_t &= \phi_1 y_{t-1} + \phi_2 y_{t-2} + \cdots + \phi_p y_{t-p} \nonumber \\
    &\quad + \epsilon_t + \theta_1 \epsilon_{t-1} + \theta_2 \epsilon_{t-2} + \cdots + \theta_q \epsilon_{t-q}
\end{align}

where $\hat{y}_t$ represents the predicted price at time $t$, and $\epsilon_t$ is the residual (error term). The prediction accuracy was computed based on the scaled absolute error between the predicted price and the observed price at time step $T$, normalized by the standard deviation of the past 80 time steps.

\textbf{EMA.} The Exponential Moving Average (EMA) is a simple yet effective smoothing technique that gives exponentially higher weights to more recent observations, allowing for more emphasis on short-term trends. The recursive update for EMA is given by:

\begin{align}
    \hat{p}_{t+1} = \alpha p_t + (1 - \alpha) \hat{p}_t
\end{align}

where $\alpha$ is the smoothing parameter. In this study, the initial forecast values were estimated by treating them as learnable parameters and minimizing the mean squared error between the predicted and observed prices.

\textbf{Naive Constant Prediction}

As a lower-bound baseline, we included a Naive Constant Prediction approach, where the model consistently predicts the same class (Positive, Negative, or Neutral) for all time steps. This simplistic method provides a benchmark for evaluating the improvements offered by more advanced methods. The prediction probabilities for this model were set to 1 for the predicted class and 0 for the others.


\section{Results and Discussion}

We present a detailed analysis of the experimental results focusing on the sign prediction accuracy for both the 2-class and 3-class tasks. As shown in Table 1, the CTTS model consistently outperforms all baseline methods, including DeepAR, ARIMA, and EMA. For the 2-class task, random guessing yields a 50\% accuracy baseline, while for the 3-class task, it results in a 33\% baseline. However, CTTS significantly surpasses these baselines, with substantial improvements across both tasks. This confirms the efficacy of combining CNNs and Transformers for time series forecasting, as it enhances the model's ability to capture intricate temporal dependencies across varying time scales.

Additionally, the results highlight that CTTS provides a notable improvement over baselines in scenarios where both short-term and long-term dependencies are present. This is particularly evident when comparing CTTS with traditional models like ARIMA and EMA, which primarily focus on linear temporal patterns and are less adept at handling complex, non-linear dependencies. The superior performance of CTTS demonstrates the model’s flexibility in adapting to diverse market dynamics and capturing multi-horizon relationships in the data.

Furthermore, we introduce a thresholded variant of the accuracy metric, wherein we evaluate only those predictions whose associated class probability exceeds the 75th percentile. This strategy allows us to focus on the high-confidence predictions and assess the reliability of the models under more stringent conditions. As shown in the “2-class*” and “3-class*” columns of Table 1, the thresholding technique leads to increased accuracy for all evaluated methods. However, the improvement is most pronounced for CTTS, where the gain in accuracy is the highest. This indicates that CTTS not only generates accurate predictions but also provides reliable confidence estimates, which are crucial in practical decision-making environments. 

The significant performance boost observed in the thresholded 3-class task further underscores CTTS's robustness. The ability of CTTS to maintain high performance in scenarios that require high-confidence predictions illustrates its potential for real-world applications, particularly in fields like financial trading where precision is paramount. Specifically, CTTS's accurate class probability predictions enable the model to filter out erroneous low-confidence predictions, thus enhancing the overall predictive reliability. In high-stakes domains such as stock trading, this capacity to deliver accurate high-confidence forecasts could have considerable economic implications. For instance, CTTS could be employed in a trading strategy where buy/sell/hold decisions are made based on the predicted class of stock movement (up, down, or flat), and the magnitude of trades could be adjusted according to the predicted probability. This ensures that high-confidence predictions translate into more aggressive trading, while low-confidence predictions are treated with caution.

Moreover, the results suggest that CTTS effectively bridges the gap between traditional statistical methods and modern deep learning-based approaches, offering the best of both worlds. Its convolutional layers enable local pattern recognition, while the transformer layers capture long-range dependencies, leading to a more comprehensive understanding of the underlying data structure. This combination not only improves forecasting performance but also enhances the model's ability to generalize across different market regimes, which is particularly valuable for volatile financial data.

\section{Conclusion}

In this work, we tackled the problem of financial time series forecasting, particularly focusing on stock price movements. Through extensive experiments on intraday stock price data from S\&P 500 companies, we demonstrated that CTTS significantly outperforms existing methods, including ARIMA, EMA, and DeepAR. Our results show that CTTS can capture both short-term and long-term dependencies effectively, while also providing reliable probability estimates, which are crucial for high-confidence decision-making. These findings suggest that CTTS has strong potential for application in real-world financial forecasting and trading systems.

\nolinenumbers

\bibliography{library}

\bibliographystyle{abbrv}

\end{document}